\documentclass[runningheads]{llncs}
\usepackage{graphicx}
\usepackage{tikz}
\usepackage{multirow}
\usepackage{algorithm}
\usepackage{algpseudocode}
\usepackage{amssymb}
\usepackage{comment}

\begin{document}
\title{Towards a Practical Defense against Adversarial Attacks on Deep Learning-based Malware Detectors via Randomized Smoothing}
\titlerunning{Randomized Smoothing for Malware Detection}
%
\author{Daniel Gibert\inst{1} \and
Giulio Zizzo\inst{2} \and
Quan Le\inst{1}}
\authorrunning{D. Gibert et al.}
%
\institute{CeADAR, University College Dublin, Dublin, Ireland \email{\{daniel.gibert,quan.le\}@ucd.ie}\and 
IBM Research Europe, Dublin, Ireland\\
\email{giulio.zizzo2@ibm.com}\\}
\maketitle              
\begin{abstract}
Malware detectors based on deep learning (DL) have been shown to be susceptible to malware examples that have been deliberately manipulated in order to evade detection, a.k.a. adversarial malware examples. More specifically, it has been show that deep learning detectors are vulnerable to small changes on the input file. Given this vulnerability of deep learning detectors, we propose a practical defense against adversarial malware examples inspired by randomized smoothing. In our work, instead of employing Gaussian or Laplace noise when randomizing inputs, we propose a randomized ablation-based smoothing scheme that ablates a percentage of the bytes within an executable. During training, our randomized ablation-based smoothing scheme trains a base classifier based on ablated versions of the executable files. At test time, the final classification for a given input executable is taken as the class most commonly predicted by the classifier on a set of ablated versions of the original executable. To demonstrate the suitability of our approach we have empirically evaluated the proposed ablation-based model against various state-of-the-art evasion attacks on the BODMAS dataset. Results show greater robustness and generalization capabilities to adversarial malware examples in comparison to a non-smoothed classifier.

\end{abstract}
\section{Introduction}
Nowadays, machine learning (ML) is being adopted to enhance anti-malware engines as it has the capability to ``learn without being explicitly programmed". Typically, ML-based malware detectors can be divided in two groups: (1) static detectors and (2) dynamic detectors. On the one hand, static malware detectors analyze the characteristics of a program or file without executing it. They rely on examining the static attributes, such as the binary file's structure, header information, or metadata, to determine if a file is potentially malicious. These detectors operate on the assumption that certain patterns or properties found in malware can be identified without running the code. On the other hand, dynamic malware detectors analyze the behavior of a program or file by executing it in a controlled environment. These detectors observe the program's actions, interactions with the system, network traffic, and other run-time behaviors to determine if a file exhibits malicious activities such as unauthorized data access, network connections to known malicious domains, or attempts to modify system files. In this work, we are going to focus on static detectors, specifically in static ML-based malware detectors, as extracting information from an executable without executing it is faster than running an executable and logging its behavior. In addition, static ML-based detectors are usually deployed as a first line of defense, alongside with signature-based and heuristic-based approaches, in a multi-layer defense malware detection system.

There are currently two primary categories of static ML-based malware detectors: (1) Feature-based detectors and (2) end-to-end detectors. On the one hand, feature-based detectors rely on the process of feature engineering, where manually crafted features are extracted from benign and malicious samples to train machine learning models. These detectors aim to capture specific characteristics or patterns that distinguish malicious software from benign programs. The feature extraction process typically involves techniques such as n-gram analysis, entropy calculation, API information and structural analysis of the binary files. These extracted features serve as input to traditional machine learning algorithms such as gradient boosting trees~\cite{2018arXiv180404637A}, support vector machines~\cite{DBLP:conf/malware/SaxeB15} (SVM), random forests~\cite{DBLP:conf/codaspy/AhmadiUSTG16,GIBERT2022117957}, and feed-forward-neural networks~\cite{236304}. 

Feature-based detectors require a great deal of domain knowledge to extract a set of features that accurately represent the executables. This process is time consuming and requires deep understanding of the executable's file format. Consequently, recent research has focused on constructing models that can extract their own features, which are known as end-to-end detectors~\cite{DBLP:conf/ccia/GibertBMPSV17,DBLP:conf/aaai/RaffBSBCN18,krcal2018deep}. For instance, Raff et al.~\cite{DBLP:conf/aaai/RaffBSBCN18} introduced MalConv, a shallow CNN architecture that can learn features directly from raw byte inputs by performing convolutions.
The MalConv model, like any other model that directly learns from the raw byte representation of executables, is trained using both benign and malicious code, and identifies patterns shared among those executables. Malware authors are aware of this and they try to disguise their malicious code in a way that it resembles benign code, thereby causing the target malware detector to incorrectly classify the malicious executable as benign.
An effective method to evade detection by end-to-end detectors is to inject benign code within a malicious executable, either by appending it at the end of the malicious code, which is referred to as overlay append, or by adding new sections that contain the benign code~\cite{DBLP:journals/tifs/DemetrioBLRA21}. In addition, more elaborated evasion tactics have been proposed to evade end-to-end malware detectors. For instance, Demetrio et al.~\cite{demetrio2021functionality} proposed GAMMA, a genetic adversarial machine learning malware attack which optimizes the benign content injected into adversarial malware examples using a genetic algorithm. Moreover, Kreuk et al.~\cite{DBLP:journals/corr/abs-1802-04528} adapted the Fast Gradient Sign Method (FGSM) to create a small adversarial payload that flips the prediction made for malicious executables from malicious to benign. 

In light of the susceptibility of deep learning detectors to even minor manipulations in the input file, we propose a practical defense against adversarial malware examples based on randomized smoothing~\cite{DBLP:conf/icml/CohenRK19,DBLP:conf/nips/LiCWC19,DBLP:conf/sp/LecuyerAG0J19}. In this work, instead of employing Gaussian or Laplace noise when randomizing inputs, we propose a randomized ablation-based smoothing scheme that ablates a proportion of the bytes in a executable file.


The main contributions of this work are the following:
\begin{itemize}
    \item We propose the first model agnostic adversarial defense technique against adversarial malware examples, i.e. you can use any machine learning model as a base classifier.
    \item We present a randomized ablation-based smoothing classification scheme specifically designed for the task of malware detection. 
    \item We empirically evaluate the proposed randomized ablation-based smoothing scheme against state-of-the-art evasion attacks to assess its robustness compared to a non-smoothed classifier.
\end{itemize}

\section{Problem Formulation}
In this section, we present the background to the task of malware detection along with an overview of the latest evasion attacks in the literature. 

\subsection{The Task of Malware Detection}
Malware detection refers to the task of determining whether a given software program is benign or malicious. Traditional approaches rely on costly and time-consuming feature engineering, but deep learning-based detectors use one or more convolutional layers to directly learn patterns from raw bytes. 
However, similarly to other ML models, deep learning malware detection models are susceptible to adversarial examples~\cite{DBLP:journals/corr/SzegedyZSBEGF13,DBLP:conf/pkdd/BiggioCMNSLGR13}. In the context of malware detection, given a malware detector $f$, the goal of an adversarial attack on a malware example $x$ is to produce an adversarial malware example $x_{adv}$, such that $x_{adv}$ has the same functionality as $x$, but $f$ misclassifies $x_{adv}$ as a benign example.
Unlike attacks on image classifiers, the adversarial examples for malware detection do not need to be visually indistinguishable from the original example; they simply need to maintain the executable's functionality.

\subsection{Evasion Attacks}
ML-based malware detectors have been shown to be very sensitive to small changes in the input file, and can be easily bypassed by injecting carefully crafted adversarial payloads~\cite{DBLP:conf/sp/SuciuCJ19}.
Recent advanced evasion attacks introduced in the literature to evade end-to-end detectors can be broadly categorized into two groups depending on the level of access and knowledge the malware authors have of the target detector: (1) white-box attacks~\cite{DBLP:journals/corr/abs-1802-04528,DBLP:conf/sp/SuciuCJ19} and (2) black-box attacks~\cite{DBLP:journals/tifs/DemetrioBLRA21,YUSTE2022102643}. White-box attacks require complete knowledge of the detector, including training algorithm, input and output, and access to the model parameters. 
For instance, Suciu et al.~\cite{DBLP:journals/corr/abs-1802-04528,DBLP:conf/sp/SuciuCJ19} adapted the Fast Gradient Method (FGM) from~\cite{DBLP:journals/corr/GoodfellowSS14} to generate a small adversarial payload that caused malicious executables to be classified as benign. Moreover, Demetrio et al.~\cite{demetrio2021adversarial} presented various attacks that manipulate the format of Portable Executable files to inject an adversarial payload.
These attacks, named Full DOS, Extend and Shift, inject the adversarial payload by manipulating the DOS header, extending it, and shifting the content of the first section, respectively. On the other hand, black-box attacks do not require such comprehensive knowledge of the detector, and can be executed with limited information, i.e. the score (score-based attacks) or the label (label-based attacks) predicted by the malware detector. However, in a real-world scenario only label-based attacks are feasible as the malware authors will know nothing about the detection system and only the label associated to the submitted file will be available to them. For example, Demetrio et al.~\cite{demetrio2021functionality} proposed GAMMA (Genetic Adversarial Machine Learning Malware Attack), a label-based attack that relies on a genetic algorithm to select which benign code to inject and modify in order for the adversarial malware example to evade detection. Similarly, Yuste et al.~\cite{YUSTE2022102643} presented a method for generating adversarial malware examples by dynamically extending unused blocks, referred to as code caves, into malware binaries. Afterwards, a genetic algorithm is employed to optimize the content inserted into these code caves.

\subsection{Adversarial Defenses}
The only defense that has been published so far is adversarial training. In their work, K. Lucas et al.~\cite{advtrain:sec2023} used data augmentation to train an end-to-end malware classifier to be robust against three state-of-the-art evasion attacks, (1) In-Place Replacement attack (IPR)~\cite{10.1145/3433210.3453086}, (2) Displacement attack (Disp)~\cite{10.1145/3433210.3453086} and (3) Padding attack~\cite{DBLP:journals/corr/abs-1802-04528}. To this end, they augmented the training data by (1) applying unguided transformations of the same type of attacks, e.g. IPR and Disp; (2) using modified versions of IPR and Disp adversarial examples; (3) training with padded adversarial examples; using adversarial examples perturbated with attacks adapted from other discrete domains~\cite{10.5555/3455716.3455759,ijcai2019p669}. 

Although effective, adversarial training makes the malware classifier robust only to the attacks employed to augment the training data.
To this end, in our work we present a smoothing-based approach to improve the robustness of end-to-end malware classifiers. In contrast to adversarial training, smoothing-based classifiers aim to smooth out the decision boundaries in the input space, making the classifiers more robust to a wider range of attacks.

\section{Methodology}
In this section we introduce our smoothing-based defense which we adapt from the Computer Vision domain by: (1) replacing the standard Gaussian randomization scheme with a randomized ablation-based scheme that operates on the bytes of an executable; (2) introducing a training procedure that takes into account that files are variable-length byte arrays.

\subsection{Randomized Smoothing}
Smoothing is a method used in robust machine learning that averages a model's output in relation to randomized inputs.
This method has recently gained attention in the computer vision domain due to its capacity to reduce a model's sensitivity to noise or fine-scale variations. More specifically, it has been shown that randomized smoothing schemes employing Gaussian or Laplace noise when randomizing inputs, provide $l_{p}$ robustness certificates~\cite{DBLP:conf/icml/CohenRK19,DBLP:conf/nips/LiCWC19,DBLP:conf/sp/LecuyerAG0J19}. 
These randomization techniques, however, are inadequate for the task of malware detection because (1) they erroneously assume numerical input values, and (2) they erroneously assume all input examples are of the same size. However, byte values are categorical and are often embedded as a vector of real values, and the size of the input files vary.
To address these incompatibilities, we propose a randomized ablation-based (RA) smoothing scheme which randomizes inputs by randomly ablating a percentage of the bytes in a given binary file.

\subsubsection{Randomized Ablation-based Smoothing Scheme}
\label{sec:randomized_smoothing}
The set of possible byte values in an executable will be represented by $S = \{ 0, 1, ..., 255\}$. We will use $X = S^{d}$ to represent the set of possible executables, where $d$ is the maximum length of the  executables in $X$ that will be used for classification. Larger executables will be clipped while smaller executables will be padded with the specially-encoded 'PAD' symbol.

\begin{remark}
The 'PAD' token is embedded as a E-dimensional array of zeros, where E is the embedding size, to indicate the absence of information about a byte. For instance, if E=8, then the embedding of the 'PAD' token is [0,0,0,0,0,0,0,0].
\end{remark}

In our randomized ablation-based smoothing scheme, a base malware classifier, $f$, is trained to make classifications based on an ablated version $\tilde{x}$ of a given executable file $x$. This ablated version $\tilde{x}$ consists of a copy of the original executable file $x$, with all the selected
byte values replaced/ablated with the specially-encoded 'PAD' symbol. The rationale behind is to conceal the information of the selected bytes from the classifier. The training procedure is defined in Algorithm~\ref{alg:smoothed_classifier_training}.

\begin{algorithm}
\caption{Smoothed classifier training procedure}\label{alg:smoothed_classifier_training}%
\begin{algorithmic}
\Require training dataset $D_{train}$, malware detector $f$ with parameters $\theta$, probability of ablating a byte $p \in \mathbb{R} \: ,\:  0<=p<=1$
\State $\theta \gets Initialize\: parameters $
\For{i=1, MAX\_EPOCHS}
    \For{$x, y \in D_{train}$}
        \State $\tilde{x} \gets ABLATE(x, p)$
        \State $\tilde{y}   \gets f(\tilde{x} )$
        \State $Loss \gets criterion(y, \tilde{y} )$
        \State $\theta \gets Update\:  parameters$
    \EndFor
\EndFor
\end{algorithmic}
\end{algorithm}

Algorithm~\ref{alg:ablate_operation} defines the operation $ABLATE(x, p)$ which takes as input an executable file $x$, represented as an array of byte values, and a probability $p$, which denotes the probability of ablating every byte in $x$, and outputs an ablated file $\tilde{x}$, i.e. an array of byte values, with all the selected ablated bytes replaced with the 'PAD' token. Ablating the bytes given a probability $p$, instead of ablating a fixed number of bytes, allows us to ablate a number of bytes proportional to the size of the input file.

\begin{remark}
Notice that contrarily to the Computer Vision (CV) domain, where it is common to downscale all images to a fixed size, i.e. 28×28 for MNIST, 32×32 for CIFAR10, or 225 x 255 for Imagenet, the size of the files is different from one another, 97 bytes being the smallest size a Portable Executable file can have. Choosing to ablate a fixed number of bytes indistinctively of the size of the file is impractical as it might generate situations where the smallest files would have almost all bytes ablated and the bigger files would have a very small proportion of their bytes ablated.
To circumvent this limitation, we ablate a number of bytes proportional to the size of the file and we sample batches of similarly sized files during training. This allows us to minimize excess padding during training.
\end{remark}

The randomized ablation-based smoothing scheme can be defined as a two-stage process.  
In the first stage, given an input file $x$ of size $d$, we select which bytes are going to be ablated based on a probability $p$. This operation is referred as $CREATE\_MASK(x, p)$. The output of the $CREATE\_MASK(x, p)$ is an array, referred to as $m$, of size equals to $d$ consisting of 0s and 1s, indicating whether or not to ablate a particular byte. For instance, if the i-th element of $m$ equals 1 it indicates that the i-th byte in $x$ has to be ablated. For example, $CREATE\_MASK([90, 00, 03, 00, 00, 04, ..., 13], 0.40) = [0, 1, 1, 0, 0, 0,...,1]$
In the second stage, the original file is ablated according to a given mask to yield a new file, with the selected bytes ablated/replaced with the specially-encoded 'PAD' token. This operation is referred to as $APPLY\_MASK(x, m)$. For example,  $APPLY\_MASK([90, 00, 03, 00, 00, 04, ..., 13], [0,1,1,0,0,0,...,1]) = [90, \\PAD, PAD, 00, 00, 04, .., PAD]$.

\begin{algorithm}
\caption{ABLATE operation}
\begin{algorithmic} 
\Function{CREATE\_MASK}{$x,p$}
    \State m $\gets$ [0]*$|x|$
    \For{i $\gets$ 1 to  $|x|$}
        \State r $\gets$ random()
        \If{r <= p}
            \State m[i] $\gets$ 1
        \EndIf
    \EndFor
    \State \Return $m$
\EndFunction

\Function{APPLY\_MASK}{$x, m$}
    \State $\tilde{x}$ $\gets$ copy(x)
    \For{i $\gets$ 1 to  $|x|$}
        \If{m[i] == 1}
            \State $\tilde{x}$[i] $\gets$ 'PAD'
        \EndIf
    \EndFor
    \State \Return $\tilde{x}$
\EndFunction

\Require : a file, p : a probability $\in \{0,1\}$.
\State $m \gets CREATE\_MASK(x, p)$
\State $\tilde{x} \gets APPLY\_MASK(x, m)$
\State \Return $\tilde{x}$
\end{algorithmic}
\label{alg:ablate_operation}
\end{algorithm}

\begin{figure}[ht]
    \includegraphics[width=\textwidth]{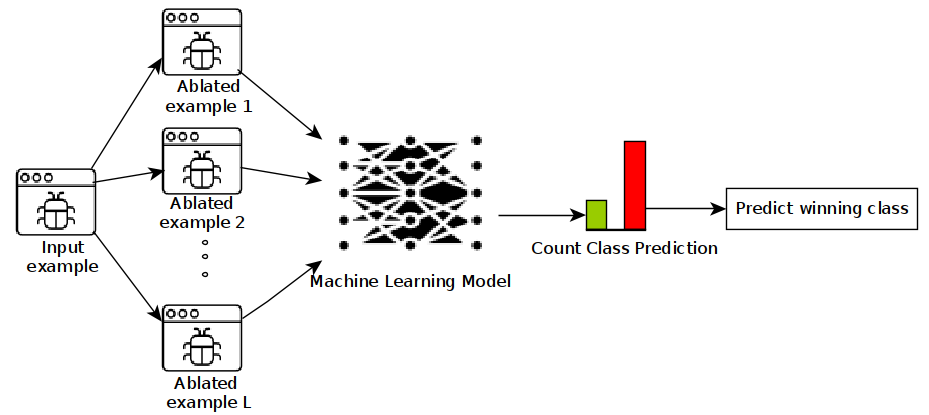}
    \centering
    \caption{Randomized ablation-based smoothing scheme overview.}
    \label{fig:randomized_scheme_overview}
\end{figure}

Let $x$ be an input file, $p$ be the probability of a given byte of the file to be ablated, and $f$ be a base classifier. At test time, we generate $L$ ablated versions of $x$ using the function $ABLATE(x,p,L)$, and classify each ablated version into its corresponding class using $f$. To make the final classification, we count the number of ablated versions of $x$ that the base classifier returns for each class and divide it by the total number of ablated versions $L$. For each class, i.e. benign or malicious, $f(\tilde{x})$ will either be 0 or 1. However, it is not required that $f(\tilde{x}) = 1$ for any class. This functionality can be implemented by a threshold. The classifier may abstain, returning zero for all classes. On the other hand, the classifier cannot return 1 for multiple classes. An overview of the scheme is represented in Figure~\ref{fig:randomized_scheme_overview}.

This problem can be formally defined as follows. Let $C = \{c_1, c_2, ..., c_k\}$ be the set of all possible classes. For each class $c_i \in C$, let $N_i$ be the number of ablated versions of $x$ that the base classifier $f$ returns as belonging to class $c_i$. 
Then, the probability of input file $x$ belonging to class $c_i$ can be estimated as:

$$P(c_i, x) = \frac{N_i}{L}$$

where $N_i$ is the number of ablated versions of $x$ that belong to class $c_i$ according to the base classifier $f$.

The function $ABLATE(x,p,L)$ generates $L$ ablated versions of input file $x$ by randomly ablating bytes from $x$ with probability $p$ of the content from $x$ and replacing it with the PAD token. The ablated versions are then returned as a set of $L$ files.

The final classification of $x$ can then be determined as:

$$\hat{y}=argmax_{c_{i}\in C}P(c_i|x)$$

where $\hat{y}$ is the predicted class for input file $x$.


\begin{remark}
    For the task of malware detection there are only two classes indicating whether the executable is benign or malicious, i.e. $c=2$. However, the approach could be easily extended to a multi-class classification problem such as the task of malware classification. Malware classification is the process of assigning a malware sample to a specific malware family. In this case, c will be equal to the number of malware families. 
\end{remark}

\section{Evaluation}
In this section we empirically evaluate the robustness of the proposed randomized ablation-based (RA) smoothing scheme against various state-of-the-art evasion attacks.

\subsection{Experimental Setup}
Following you can find the details of the experimental setup, including data sources, machine learning models for malware detection, and parameters for the randomized ablation-based smoothing scheme. The experiments have been run on a machine with an Intel Core i7-7700k CPU, 1xGeforce GTX1080Ti GPU and 64Gb
RAM. The code has been implemented with PyTorch~\cite{NEURIPS2019_9015} and will be publicly available in our Github repository~\footnote{\url{https://github.com/danielgibert/randomized\_smoothing\_for\_malware\_detection}} after publication.

\subsubsection{BODMAS Dataset}
\label{sec:bodmas_dataset}
In this paper, we use the BODMAS dataset~\cite{bodmas} to evaluate the proposed randomized ablation-based classification scheme. This dataset consists of 57,293 malware with family information (581 families) and 77,142 benign Windows PE files collected from August 2019 to September 2020. The dataset has been partitioned into three sets, training (80\%), validation (10\%) and test sets (10\%), taking into account the timestamp of each sample, i.e. examples in the training set contain the older executables while the examples in the test set contain the most recent executables. To speed-up the experiments we have only considered those executables that are equal or smaller than 1Mb. The rationale behind is that the greater the input size the greater the computational time required to run the experiments. Furthermore, as some of the evasion attacks evaluated manipulate the executables by injecting content, the bigger executables would have to be clipped to feed the model. By only using executables that are equal or smaller than 1Mb we avoid having to clip the executables and thus, losing important information for classification.
In consequence, the reduced dataset consists of 39,380 and 37,739 benign and malicious executables, respectively.

\subsubsection{Malware Detectors}
\begin{figure}[ht]
    \includegraphics[width=\textwidth]{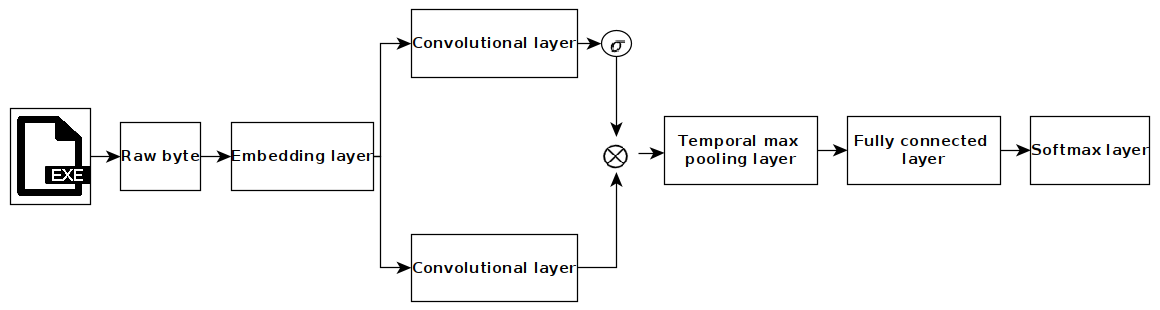}
    \centering
    \caption{Randomized ablation-based smoothing scheme overview.}
    \label{fig:malconv_architecture}
\end{figure}

In this work, we experiment with a deep learning-based malware detector called MalConv~\cite{DBLP:conf/aaai/RaffBSBCN18}. MalConv is one of the first end-to-end deep learning model proposed for malware detection. End-to-end models learn to classify examples directly from raw byte sequences, instead of relying on manually feature engineering. The network architecture of MalConv consists of an embedding layer, a gated convolutional layer, a global-max pooling layer and a fully-connected layer. See Figure~\ref{fig:malconv_architecture}.

Using MalConv as a basis, two malware detectors have been evaluated:

\begin{itemize}
    \item NS-MalConv. This detector corresponds to a non-smoothed MalConv model. It serves as a non-robust baseline as no specific technique has been employed to improve robustness to evasion attacks. 
    \item RA-MalConv. This detector implements the randomized ablation-based smoothing scheme proposed in Section~\ref{sec:randomized_smoothing} using MalConv as a base detector.
\end{itemize}

\subsection{Empirical Evaluation}
In this section, we empirically evaluate the robustness of the randomized ablation-based smoothing scheme to several published evasion attacks. By doing so, we aim to provide a complete picture of the strengths and weaknesses of the proposed defense. First, we provide details of the performance of the proposed defense on non-adversarial examples in Section~\ref{sec:nonadversarial_evaluation}. Afterwards, we examine the robustness of the proposed ablation-based scheme against three state-of-the-art evasion attacks in Section~\ref{sec:sota_attacks}.

\subsubsection{Non-adversarial Evaluation}
\label{sec:nonadversarial_evaluation}
The randomized-based ablation scheme is controlled by two parameters: (1) the proportion of bytes to ablate with respect of the file size, denoted by $p \in [0,1]$, and the number of ablated versions of a given file that are used for classification, denoted by $L$. These two parameters serve as input to the methods and affect the way the method operates and the output it produces. Following we analyze the performance of the RA-MalConv model when using different various values for $p \in \{0.10, 0.20, 0.30, 0.40, 0.50\}$ in Table~\ref{tab:sa_malconv_proportion_bytes}. In our experiments, we generate 100 ablated examples for each example that the RA-MalConv model has to classify ($L=100$). This is done to make the computational cost of the defense more manageable. As observed in Table~\ref{tab:sa_malconv_proportion_bytes}, the highest accuracy and F1-score on the validation set is achieved when $p=0.20$. For this reason, For the rest of the experiments we use $p=0.20$ (20\% chance of ablating a byte) for the RA-MalConv model.

Table~\ref{tab:non_adversarial_examples_performance_metrics} presents the performance metrics of the NS-MalConv and RA-MalConv models on the validation and test sets. Results provide evidence that the RA-MalConv model has similar detection accuracy and F1-score compared to the NS-MalConv on clean data. Next, we evaluate our randomized ablation-based Malconv model against various evasion attacks to assess its robustness.


\begin{table}[ht]
\centering
\caption{Performance metrics of the RA-MalConv for different proportions of bytes ablated.}
\label{tab:sa_malconv_proportion_bytes}
\begin{tabular}{l|lllll}
\hline
\multirow{2}{*}{RA-MalConv} & \multicolumn{5}{l}{Probability of ablating a byte ($p$)}                                                                                            \\ \cline{2-6} 
                             & \multicolumn{1}{l|}{0.10}  & \multicolumn{1}{l|}{0.20}  & \multicolumn{1}{l|}{0.30}  & \multicolumn{1}{l|}{0.40}  & \multicolumn{1}{l}{0.50}  \\ \hline
Accuracy                     & \multicolumn{1}{l|}{98.04} & \multicolumn{1}{l|}{\textbf{98.08}} & \multicolumn{1}{l|}{97.43} & \multicolumn{1}{l|}{97.76} & \multicolumn{1}{l}{97.94} \\ \hline
F1-Score                     & \multicolumn{1}{l|}{98.03} & \multicolumn{1}{l|}{\textbf{98.06}} & \multicolumn{1}{l|}{97.46} & \multicolumn{1}{l|}{97.72} & \multicolumn{1}{l}{97.92} \\ \hline
\end{tabular}%
\end{table}

\begin{table*}[ht]
\centering
\caption{Performance metrics of the non-smoothed and smoothed models on the validation set, the test set and the sub-tests of 500 and 100 malware examples used for adversarial attack evaluation. With the sub-test sets being composed of only malware we simply report the accuracy.}
\label{tab:non_adversarial_examples_performance_metrics}
\begin{tabular}{ccccccc}
\hline
\multicolumn{1}{c}{\multirow{2}{*}{}} & \multicolumn{2}{c}{Validation set}                           & \multicolumn{2}{c}{Test set}                                 & \begin{tabular}[c]{@{}c@{}}Test sub-set\\ (500 examples)\end{tabular} & \begin{tabular}[c]{@{}l@{}}Test sub-set\\ (100 examples)\end{tabular} \\ \cline{2-7} 
\multicolumn{1}{c}{}                  & Accuracy                        & F1-score                   & Accuracy                        & F1-Score                   & Accuracy                                                              & Accuracy                                                              \\ \hline
\multicolumn{1}{l|}{NS-MalConv}       & 97.55                           & \multicolumn{1}{c|}{97.52} & \textbf{98.03} & \multicolumn{1}{c|}{97.98} & \multicolumn{1}{c|}{\textbf{97.00}}                  &      \textbf{99.00}                                                                 \\ \hline
\multicolumn{1}{l|}{RA-MalConv}       & \textbf{98.08} & \multicolumn{1}{c|}{\textbf{98.03}} & 97.61                           & \multicolumn{1}{c|}{97.57} & \multicolumn{1}{c|}{95.00}                                            &     \textbf{99.00}                                                                  \\ \hline
\end{tabular}
\end{table*}

\subsubsection{Empirical Robustness Evaluation Against SOTA Evasion Attacks}
\label{sec:sota_attacks}

We consider three recently published attacks designed to bypass static PE malware detectors as summarized in Table~\ref{tab:sota_attacks}. These attacks manipulate the executables by creating new spaces within the executables~\cite{demetrio2021adversarial,YUSTE2022102643} and by injecting new content in a newly-created section~\cite{demetrio2021functionality}. The maximum size of the adversarial malware examples was constrained to 2,000,000 bytes in our experiments (twice the maximum size of the examples used for training the models). Early termination is implemented, causing the attack to terminate immediately when the malware detector's prediction shifts from malicious to benign. Since some attacks might take hours to run per file, we use two smaller-sized test sets containing 500 and 100 malware examples randomly subsampled from the test set.
The test sub-set consisting of 500 examples have been employed to evaluate the malware detectors against the Shift attack~\cite{demetrio2021adversarial}, and the GAMMA attack~\cite{demetrio2021functionality}. On the other hand, the test sub-set of 100 examples has been used to evaluate the code caves optimization attack~\cite{YUSTE2022102643}. By employing a smaller subset, we aim to reduce the computational overhead and accelerate the overall experimentation process~\footnote{We would like to denote that our evaluation set is comparable in size to prior work~\cite{DBLP:conf/eusipco/KolosnjajiDBMGE18,DBLP:journals/corr/abs-1802-04528,DBLP:conf/sp/SuciuCJ19,demetrio2021adversarial}}. In addition, we reduce from 100 to 20 the number of ablated versions generated for each input example, i.e. $L=20$.

\begin{table}[ht]
\centering
\caption{Description of the evasion attacks used to assess the robustness of the proposed smoothed classier.}
\label{tab:sota_attacks}
\resizebox{\textwidth}{!}{%
\begin{tabular}{l|l|l|l}
\hline
\textbf{Attack}                                                                & \textbf{Type of Attack} & \textbf{Optimizer}            & \textbf{Description}                                                                                        \\ \hline
Shift~\cite{demetrio2021adversarial}                                                                 & White-box      & Single gradient step & Shift section content                                                                              \\ \hline
GAMMA~\cite{demetrio2021functionality}                                                                 & Black-box      & Genetic algorithm    & \begin{tabular}[c]{@{}l@{}}Padding and injection of \\ benign sections\end{tabular}                \\ \hline
\begin{tabular}[c]{@{}l@{}}Optimization of \\ code caves~\cite{YUSTE2022102643}\end{tabular} & Black-box      & Genetic algorithm    & \begin{tabular}[c]{@{}l@{}}Dynamically introducing code \\ caves within an executable\end{tabular} \\ \hline
\end{tabular}
}
\end{table}

We would like to point out that the white-box attack cannot be directly applied to our randomized ablation-based classifier as it requires computing the gradients and our approach makes computing the gradients a difficult task. To circumvent this situation, we optimized the adversarial payload using genetic algorithms (GAs). The use of genetic algorithms to optimize the adversarial payload allows us to convert the white-box attack to a black-box attack. Therefore, there is no need to access the implementation details of the model or the gradient information to optimize the adversarial payload. Instead, the GA will explore the search space by imitating the process of evolution through various bioinspired operators (selection, crossover, and mutation) to optimize an initial population of solutions. Similarly to GAMMA~\cite{demetrio2021functionality}, the initial population of solutions is initialized with benign content. In our experiments, the size of the population is 50. Afterwards, the GA iteratively applies selection, crossover and mutation to optimize the adversarial payload. The implementation of each of the bioinspired operators within the GA is based on the results of Yuste et al.~\cite{YUSTE2022102643}.
\begin{enumerate}
    \item Selection operator. The selection operator selects a subgroup of 10 individuals within the current population to be crossed in the following step. This individuals are selected using a combination of elitism (only the best individuals are chosen) and a selection based on tournaments (each tournament consists of selecting ten individuals at random and the best among them is selected for the next iteration). 
    \item Crossover operator. The crossover operator pairs of selected individuals (parents) are combined to produce offspring. Each individual from the elitist group is crossed over with each individual from the tournament group, generating two offsprings per crossover. The first chunk of each offspring is the first chunk of one of the parents, the second chunk of each offspring is the second chunk from the other parent, and so on.
    \item Mutation operator. The mutation operator mutates the offspring individuals obtained after the crossover operator with probability $p_{1} = 0.1$. Then, a percentage of the genes of the individual selected for mutation are mutated with probability $p_{2} = 0.1$.
\end{enumerate}

\begin{figure}[ht]
    \includegraphics[width=\textwidth]{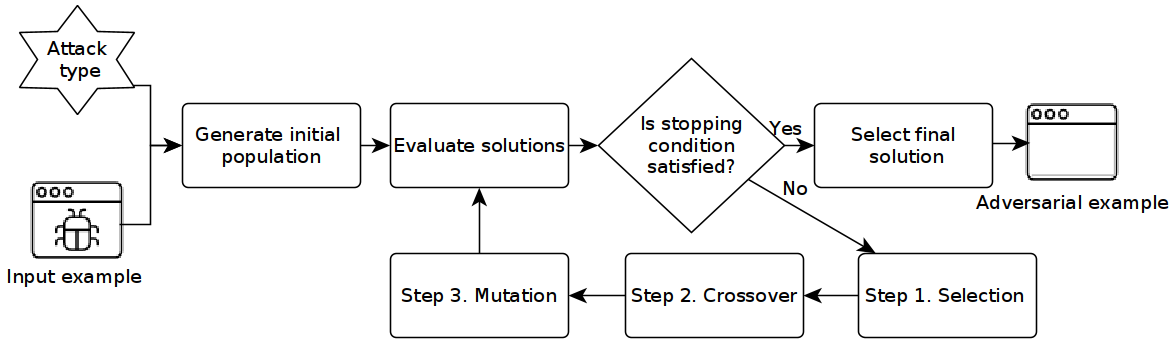}
    \centering
    \caption{Genetic algorithm scheme for malware evasion attacks.}
    \label{fig:optimization_of_bytes_with_GA}
\end{figure}

Figure~\ref{fig:optimization_of_bytes_with_GA} depicts the general scheme of the GA to the problem of malware evasion. In our case, the attack type is the shift attack although the genetic algorithm could be used to optimize any adversarial payload.

Table~\ref{ref:empirical_evaluation_sota_attacks} present the detection accuracy of the NS-MalConv and RA-MalConv models on the adversarial malware examples generated by the evasion attacks based on genetic algorithms described in Table~\ref{tab:sota_attacks}. The table shows that, regardless of the evasion attack, the RA-MalConv outperforms NS-MalConv by some margin.
For instance, NS-MalConv detects 43.40\% and 40.80\% of the adversarial malware examples generated by the Shift attack while the RA-MalConv detects 69.40\% and 67.00\% of the adversarial examples, respectively. Furthermore, RA-MalConv provides greater detection accuracy than NS-MalConv on the adversarial malware examples generated by the GAMMA attack, showing that the proposed defense mechanism holds its ground against evasion attacks that inject and optimize great amounts of benign content. Lastly, the RA-MalConv detects 81\% of the examples compared to only a 14\% detection rate of the NS-MalConv when the adversariale examples have been created by extending and optimizing code caves in the executables. Overall, the robustness of the RA-MalConv to attacks that manipulate the executables is positive and a great improvement over the baseline, NS-MalConv, demonstrating that anti-malware engines can benefit from smoothing-based classifiers.

\begin{table}[ht]
\caption{Detection accuracy of the end-to-end malware detectors on the adversarial examples generated with the evasion attacks.}
\label{ref:empirical_evaluation_sota_attacks}
\begin{tabular}{l|l|cc}
\hline
\multirow{2}{*}{\textbf{Attack}}                                                              & \multirow{2}{*}{\textbf{Hyperparameters}}       & \multicolumn{2}{c}{\textbf{Detection Rate}}     \\ \cline{3-4} 
                                                                                     &                                        & \multicolumn{1}{c|}{NS-MalConv} & RA-MalConv \\ \hline
\multirow{3}{*}{Shift attack}                                                        & Preferable extension amount = 1048     & \multicolumn{1}{c|}{52.20}      & \textbf{74.00}      \\
                                                                                     & Preferable extension amount = 2048     & \multicolumn{1}{c|}{43.40}      & \textbf{69.40}      \\ 

                                                                                     & Preferable extension amount = 4096     & \multicolumn{1}{c|}{40.80}      & \textbf{67.00}      \\ \hline
\multirow{3}{*}{\begin{tabular}[c]{@{}l@{}}GAMMA attack\\ (soft label)\end{tabular}} & \#sections = 10,  population size = 10 & \multicolumn{1}{c|}{42.20}      & \textbf{66.20}      \\
                                                                                     & \#sections = 100, population size = 10 & \multicolumn{1}{c|}{56.20}      & \textbf{71.80}      \\
                                                                                     & \#sections = 100, population size = 20 & \multicolumn{1}{c|}{36.20}      & \textbf{78.80}      \\ \hline
\begin{tabular}[c]{@{}l@{}}Optimization of\\ code caves*\end{tabular}                 & Default                                & \multicolumn{1}{c|}{14.00}      & \textbf{81.00}      \\ \hline
\end{tabular}
\end{table}

\section{Discussion}
In this section we discuss the advantages and limitations of the non-smoothed and smoothed detectors, NS-MalConv and RA-MalConv, respectively.

The principal advantage of training a malware detector on the whole binary information is that it provides a comprehensive and complete representation of the input example. This can lead to overfiting to the training data, as the model may become overly sensitive to specific noise patterns in the training data and fail to generalize well to new, unseen data. Our ablation-based smoothing scheme, on the other hand, can be seen as an approach to regularize the model and prevent overfitting by introducing noise into the training data via the ablation of bytes. By adding noise to the input data, the ML model learns to ignore small variations in the data that are not relevant to the classification task. 

At test time, however, the non-smoothed detector makes predictions faster and with less computational resources than the smoothed detectors as they only assess once the maliciousness of a given input example. On the contrary, the smoothed detector is slower because it needs to assess each ablated version of the input example independently and then aggregate the predictions as shown in Table~\ref{tab:models_computational_time}. Nevertheless, the benefits of the increased generalization abilities and robustness to adversarial attacks provided by the smoothed detectors outweight the added computational cost and time required for classification.

\begin{table}[ht]
\centering
\caption{Training and testing time comparison between the smoothed and non-smoothed detectors.}
\label{tab:models_computational_time}
\begin{tabular}{c|cc}
\hline
\multirow{2}{*}{\textbf{Models}} & \multicolumn{2}{c}{\textbf{Computational Time}}                                                                                                                                 \\ \cline{2-3} 
                        & \multicolumn{1}{c|}{\begin{tabular}[c]{@{}l@{}}Training Time \\ (minutes/epoch)\end{tabular}} & \begin{tabular}[c]{@{}l@{}}Test Time \\ (seconds/example)\end{tabular} \\ \hline
NS-MalConv & \multicolumn{1}{c|}{22.06} & 0.0161\\
RA-MalConv & \multicolumn{1}{c|}{44.95} & 3.3814\\\hline
\end{tabular}%
\end{table}

\section{Conclusions}
In this paper, we present the first model agnostic adversarial defense against adversarial malware examples. Building upon prior research on randomized smoothing, we introduce a randomized ablation-based smoothing scheme to build robust static end-to-end learning-based classifiers. To the best of our knowledge, this is the first time randomized smoothing has been implemented for end-to-end malware detection, i.e. taking as input the whole binary as a sequence of bytes. The novel application of our randomized ablation-based smoothing scheme creates a new robust model that generalizes better than the non-smoothed MalConv on adversarial malware examples, achieving a significant advancement in the field.

\subsection{Future Work}
Our results suggest a number of directions for future work. The most apparent direction identified is the incorporation of recent defenses from the Computer Vision domain other than randomized smoothing. It would be interesting to adapt de-randomized smoothing into static malware detection models to prevent a localized injection of benign code or adversarial content from dominating the prediction. Another line of research could be investigating approaches to identify and remove the adversarial content from the executables.

\section*{Acknowledgements}
This project has received funding from Enterprise Ireland and the European Union’s Horizon 2020 Research and Innovation Programme under the Marie Skłodowska-Curie grant agreement No 847402. The views and conclusions contained in this document are those of the authors and should not be interpreted as representing the official policies, either expressed or implied, of CeADAR, University College Dublin, and IBM Ireland Limited. We would like to thank Cormac Doherty and UCD's Centre for Cybersecurity and Cybercrime Investigation for their support.

\section*{Data and Code Availability}
The BODMAS dataset is available to the public and the source code of our approach will be available under the MIT License after the paper has been accepted for publication.

%
%
%

\bibliographystyle{splncs04}
\bibliography{refs.bib}
\end{document}